\documentclass[letterpaper, 10 pt, conference]{ieeeconf}  
\usepackage{cite}
\usepackage{amsmath,amssymb,amsfonts}
\usepackage{algorithmic}
\usepackage{graphicx}
\usepackage{textcomp}
\usepackage{multirow}
\usepackage[flushleft]{threeparttable}
\usepackage{xcolor}
\usepackage[hidelinks]{hyperref}

\IEEEoverridecommandlockouts                              

\title{\LARGE \bf
	Feature Augmented Hybrid CNN for Stress Recognition \\Using Wrist-based Photoplethysmography Sensor}

\author{Nafiul Rashid$^{*}$,~\IEEEmembership{Student~Member,~IEEE,} Luke Chen, Manik Dautta, Abel Jimenez, \\Peter Tseng, and Mohammad Abdullah Al Faruque,~\IEEEmembership{Senior~Member,~IEEE}
	\thanks{All the authors are with the Department of Electrical Engineering
		and Computer Science, University of California, Irvine, CA
		92697, USA, { $^*$Corresponding author : Nafiul Rashid (\tt\small nafiulr@uci.edu}.) \newline{\footnotesize 978-1-6654-4407-1/21/\$31.00~\copyright~2021 IEEE\hfill}}}%

\renewcommand{\baselinestretch}{.96}

\begin{document}

	
	\maketitle

\begin{abstract}
Stress is a physiological state that hampers mental health and has serious consequences to physical health. Moreover, the COVID-19 pandemic has increased stress levels among people across the globe. Therefore, continuous monitoring and detection of stress are necessary. The recent advances in wearable devices have allowed the monitoring of several physiological signals related to stress. Among them, wrist-worn wearable devices like smartwatches are most popular due to their convenient usage. And the photoplethysmography (PPG) sensor is the most prevalent sensor in almost all consumer-grade wrist-worn smartwatches. Therefore, this paper focuses on using a wrist-based PPG sensor that collects Blood Volume Pulse (BVP) signals to detect stress which may be applicable for consumer-grade wristwatches. Moreover, state-of-the-art works have used either classical machine learning algorithms to detect stress using hand-crafted features or have used deep learning algorithms like Convolutional Neural Network (CNN) which automatically extracts features. This paper proposes a novel hybrid CNN (H-CNN) classifier that uses both the hand-crafted features and the automatically extracted features by CNN to detect stress using the BVP signal. Evaluation on the benchmark WESAD dataset shows that, for 3-class classification (Baseline vs. Stress vs. Amusement), our proposed H-CNN outperforms traditional classifiers and normal CNN by $\approx$5\% and $\approx$7\% accuracy, and $\approx$10\% and $\approx$7\% macro F1 score, respectively. Also for 2-class classification (Stress vs. Non-stress), our proposed H-CNN outperforms traditional classifiers and normal CNN by $\approx$3\% and $\approx$5\% accuracy, and $\approx$3\% and $\approx$7\% macro F1 score, respectively.
	\end{abstract}
	
	\section{Introduction}

Stress is a physiological state that triggers the fight-or-flight response \cite{FoF} through chemical or hormone surge when someone perceives a new challenge or any adversarial situation. Depending on the types of challenges the stress can be - \textbf{Physical} (workout, running); \textbf{Cognitive} (solving problems, thinking); and \textbf{Emotional} (nervousness, fear, anxiety, frustration, sadness). According to American Psychological Association (APA), stress can be of 3 types based on the frequency of experiencing it \cite{APA_Stress}. \textbf{Acute} stress is a common form of stress that everyone faces for the short term. 
	Therefore, it is common to experience it and not always harmful. \textbf{Episodic acute} stress happens when someone feels stressed in a repetitive manner which happens mostly due to cognitive or emotional stress. Sometimes cognitive stress such as overworking regularly may lead to emotional stress such as anxiety, fatigue - causing episodic acute stress. British Health and Safety Executive (HSE) reports that stress, depression, or anxiety accounted for 51\% of all work-related ill health cases \cite{HSE}.
	Finally, \textbf{Chronic} stress where an individual suffers for many months or years is the major cause of clinical depression, sleep deprivation/oversleeping, abnormal body weight changes, cardiovascular diseases, or even suicide.
	It happens mostly due to emotional stress which often remains unrecognized or people deny to acknowledge it due to social stigma. Therefore, emotional stress recognition is really important as it not only hampers mental health but also has severe consequences to physical health.
	
	Moreover, this recent COVID-19 pandemic that has caused more than 4.16 Million global death as of July 2021, has increased the emotional stress among people. American Psychological Association (APA) issued a warning about the impact of these stressful events on long-term physical and mental health calling it as \textit{`A National Mental Health Crisis'} in their October 2020 report \cite{APA_Stress_2020}. Another survey on 3013 adults, released by APA on March 2021 states that - 61\% experienced undesired weight changes, 67\% had overslept, 48\% parents had increased stress, 25\% of the essential workers encountered mental health disorder and required emotional support since the start of the pandemic \cite{APA_Stress_2021}. The aforementioned facts prove that recognition of emotional stress has become more crucial now than ever.
	\section{Related Works}
	\label{related_works}
	
Recent advances in technology \cite{EMBC_2020,IoTJ_2020} have enabled the collection of stress and emotion-related physiological signals through various modalities like video, audio, and physiological sensors. Besides the advances in data analysis techniques have enabled the use of various machine or deep learning algorithms to classify or detect those states. Authors in \cite{audio_1, audio_2} used audio and/or visual data to classify different emotional states. However, such modalities are intrusive in nature and raise privacy concerns for the users. Therefore, the use of physiological signals collected through various wearable devices has been gaining momentum in stress and emotion detection.
	
Authors in \cite{cStress} use chest-worn device that captures physiological signals from Electrocardiogram (ECG), respiration (RESP) and 3-axis Accelerometer (ACC) to detect stress. Another group in \cite{wrist_worn} used a wrist-worn device recording BVP, Electrodermal Activity (EDA), Skin Temperature (TEMP), and ACC to detect stress. Researchers in \cite{music_stress} used ECG, RESP, EDA, and Electromyogram (EMG) data to detect emotions in response to music.

The datasets used in the above works are collected in-house and are not publicly available. On the other hand, authors in \cite{driver_stress} published a dataset that has ECG, EDA, RESP, and EMG data for drivers' stress detection. Another group in \cite{picard_emotion} published a dataset containing EMG, BVP, EDA, RESP signals for 8 different emotional stimuli from a single subject. Authors in \cite{DEAP} published a dataset for emotion analysis using Electroencephalogram (EEG), facial videos, and physiological signals. The aforementioned works either focused on detecting stress or emotion using wearable devices. Authors in \cite{WESAD} tried to bridge that gap by creating WESAD (Wearable Stress and Affect Detection) dataset that contains stress and emotion data using chest-worn and wrist-worn devices. They also provided a comparative analysis of individual physiological signals from chest and wrist in detecting stress using classical machine learning algorithms - Decision Tree (DT), Random Forest (RF), etc. The same authors also used the wrist-worn device in \cite{CNN_pitfall} to detect stress and emotion in the wild. Researchers in \cite{sirat_DCOSS} used the WESAD dataset to propose a sensor translation mechanism to create chest-based features from the wrist data to detect stress using classical machine learning algorithms. 

	\section{Problem Statement and Contributions}
	
	The aforementioned works in Section \ref{related_works} used multimodal wearable sensor data from either chest/wrist-worn devices to detect emotion or stress. However, wrist-worn devices are more convenient for daily use than chest ones. Besides, the wrist-worn devices used in the literature are mostly research-grade and have multiple sensors like PPG, EDA, TEMP, ACC. Among all these wrist-based sensors, PPG is mostly available in all consumer-grade wristwatches and has proven to be a strong biomarker for detecting stress \cite{WESAD}. Therefore, this paper focuses on detecting stress using a wrist-based PPG sensor suitable for daily monitoring via consumer-grade wristwatches. Moreover, state-of-the-art works have used either classical machine learning algorithms to detect stress or emotion using hand-crafted features or they have used deep learning algorithms like Convolutional Neural Network (CNN) which automatically extracts features. In this paper, we propose a novel hybrid CNN (H-CNN) that uses both the hand-crafted features and automatically extracted features by CNN to detect stress. 
	Finally, we demonstrate the effectiveness of our hybrid approach using wrist-based BVP signal from the WESAD \cite{WESAD} dataset.
	The novel contributions of this paper are as follows:
	\begin{itemize}
		\item Propose a novel hybrid CNN (H-CNN) classifier for stress detection using wrist-based PPG sensor. It uses both handcrafted features and automatically extracted features by CNN to detect stress.
		\item Validation of our proposed approach using BVP signal from WESAD \cite{WESAD} dataset collected through wrist-based PPG.
		
		\item Evaluation on the benchmark WESAD dataset shows that, for 3-class classification (Baseline vs. Stress vs. Amusement), our proposed H-CNN outperforms traditional classifiers and normal CNN by $\approx$5\% and $\approx$7\% accuracy, and $\approx$10\% and $\approx$7\% macro F1 score, respectively. Also for 2-class classification (Stress vs. Non-stress), our proposed H-CNN outperforms traditional classifiers and normal CNN by $\approx$3\% and $\approx$5\% accuracy, and $\approx$3\% and $\approx$7\% macro F1 score, respectively.
	\end{itemize}
	
	\section{Our Methodology}

	\begin{figure*}[t]
	\centering
	\includegraphics[trim={15cm 9.2cm 15.2cm 9.1cm},clip, width=\linewidth]{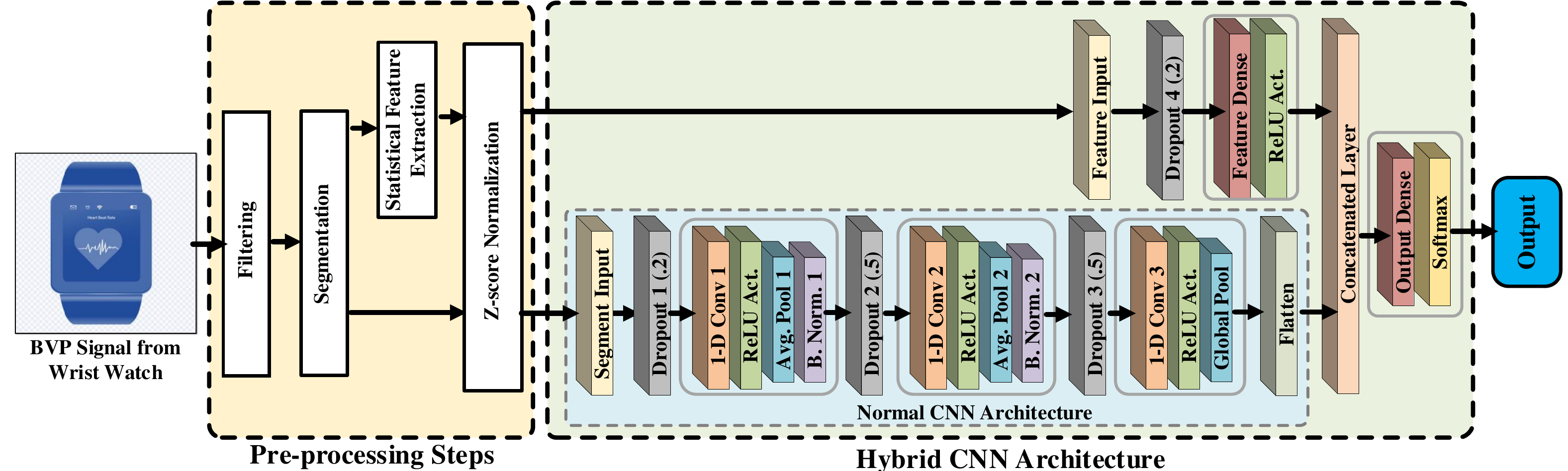}
	\caption{Overview of Our Proposed Methodology}
	\vspace{-5mm}
	\label{Methodology}
\end{figure*} 
	
	\subsection{Pre-processing Steps}
	\subsubsection{Filtering}
	As shown in Figure \ref{Methodology}, the pre-processing steps start with filtering the raw BVP signal. We filter the raw BVP signal by a butter-worth bandpass filter of order 3 with cutoff frequencies ($f_1$=.7 Hz and $f_2$=3.7 Hz). We take into account the heart rate at rest ($\approx$40 BPM) or high heart rate due to exercise scenarios or tachycardia ($\approx$220 BPM) following the method mentioned in \cite{HR_PPG}.
	\subsubsection{Segmentation}
	The filtered signal is segmented by a window of 60 seconds of data following the paper that introduced the
WESAD dataset \cite{WESAD}. We use a sliding length of 5 seconds in between the segments. Each segment contains 3840 samples as the sampling rate of the BVP signal is 64 Hz.
	\subsubsection{Feature extraction}
	The first step of the feature extraction is the detection of heartbeats. Once the peaks are detected, different time domain and frequency domain features are extracted based on the location of the peaks.  We extract the time and frequency domain features as in \cite{WESAD} to ensure a fair comparison of our H-CNN classifier against the traditional machine learning classifiers used in the WESAD paper. We use the same frequency bands - ultra-low (ULF: 0.01-0.04 Hz), low (LF: 0.04-0.15 Hz), high (HF: 0.15-0.4 Hz) and ultra-high (UHF: 0.4-1.0 Hz) band as in \cite{WESAD} to calculate different frequency domain features. The list of extracted features is given in Table \ref{extracted_features}.
	
   	\begin{table}[t]
   	\centering
   	\begin{threeparttable}
   		\caption{List of Extracted Features}
   		\label{extracted_features}
   		\begin{tabular}{|c|c|}
   				\hline
   				\textbf{Feature Symbol} & \textbf{Feature Names}  \\
   				\hline\hline
   				$\mu_{HR}$, $\sigma_{HR}$ & Mean and Standard Devaiation of HR \\
   				\hline
   				$\mu_{HRV}$, $\sigma_{HRV}$ & Mean and Standard Devaiation of HRV \\
   				\hline
   				\multirow{2}{*}{$NN50$, $pNN50$} & Number and percentage of HRV \\
   				&intervals differing more than 50 ms  \\
   				\hline
   				$rms_{HRV}$ & Root mean square of the HRV\\
   				\hline
   				$f^{x}_{HRV}$& Energy in different\\
   				$x\in{ULF, LF, HF, UHF}$& frequency component of the HRV\\
   				\hline
   				$f^{LF/HF}_{HRV}$ & Ratio of LF and HF component\\
   				\hline
   				$\sum_{x}^{f}$ & $\sum$ of the frequnecy components\\
   				$x\in{ULF, LF, HF, UHF}$& in ULF-HF\\
   				\hline
   				$rel_{x}^{f}$ & Relative power of freq. components\\
   				\hline
   				$LF_{norm}$, $HF_{norm}$ & Normalised LF and HF
component \\
   				\hline
   			\end{tabular}
  			\begin{tablenotes}
			\item \textit{Heart Rate (HR), Heart Rate Variability (HRV)}
			\end{tablenotes}
   		\end{threeparttable}
   	\vspace{-5mm}
   	\end{table}

	\subsubsection{Z-score normalization}
	Z-score normalization is performed before passing the segments and extracted features to the H-CNN architecture.
	\subsection{Hybrid CNN (H-CNN) Architecture}
	The normalized BVP segments and the corresponding features for each segment are passed to our H-CNN architecture as shown in Figure \ref{Methodology}.
	The H-CNN architecture has two input layers- Segment and feature input. The segment input layer is followed by a dropout layer (with a 20\% dropout rate) which is then followed by 3 convolution blocks. The first and second convolution blocks have - convolution, \textit{ReLU} activation, average pooling, and batch normalization layers. Both first and second convolution block is followed by dropout layers with 50\% dropout rate which are added to reduce overfitting. 
	The third convolution block has one convolution layer followed by a global average pooling layer which is also used to reduce the overfitting of the CNN. 
	For the normal CNN architecture, the output of the global average pooling layer is directly fed to the output dense layer followed by a \textit{Softmax} activation. 
	However, for the H-CNN architecture, the output of the global average pooling layer is concatenated with the feature dense layer. 
	Finally, the concatenated layer is fed to the output dense layer that is followed by the \textit{Softmax} activation.
	The details of our H-CNN architecture are shown in Table \ref{HCNN_Architecture_Details}. As shown in Table \ref{HCNN_Architecture_Details}, the total number of parameters required to classify a segment is 6846+(13*$n_c$), where $n_c$ is the number of output classes. In this paper, we perform both 2-class (Stress vs. Non-stress) and 3-class (Baseline vs. Stress vs. Amusement) classification from the WESAD dataset.

	
	\begin{table}[t]
		\centering
		\begin{threeparttable}
			\caption{Hybrid CNN Architecture Details}
			\label{HCNN_Architecture_Details}
			\begin{tabular}{|c|c|c|c|c|c|}
				\hline
				\textbf{Layer} & \textbf{Kernel} & \textbf{Stride} &  \textbf{Act.} & \textbf{Output} &  \textbf{\# of} \\
				\textbf{Name} & \textbf{Size} & \textbf{Size} &  \textbf{Func.}& \textbf{Shape} &  \textbf{Param.} \\
				\hline\hline
				Seg. Inp. & - & - & - & 3840x1 & 0\\
				\hline
				D.O. 1 & - & - & - & 3840x1 & 0\\
				\hline
				Conv 1 & 64 & 4 & ReLU & 945x8 & 520\\
				\hline
				Pool 1 & 4 & 4 & - & 236x8 & 0\\
				\hline
				B.N. 1  & - & - & - & 236x8 & 32\\
				\hline
				D.O. 2 & - & - & - & 236x8 & 0\\
				\hline
				Conv 2 & 32 & 2 & ReLU & 103x16 & 4112\\
				\hline
				Pool 2 & 4 & 4 & - & 25x16 & 0\\
				\hline
				B.N. 2 & - & - & - & 25x16 & 64\\
				\hline
				D.O. 3 & - & - & - & 25x16 & 0\\
				\hline
				Conv 3 & 16 & 1 & ReLU & 10x8 &2056\\
				\hline
				G. Pool & 4 & 4 & - & 8 & 0\\
				\hline
				Flatten & - & - & - & 8 & 0 \\
				\hline
				Feat. Inp. & - & - & - & 19 & 0\\
				\hline
				D.O. 4 & - & - & - & 19 & 0\\
				\hline
				Feat. Den. & - & - & ReLU & 4 & 80 \\
				\hline
				Concate & - & - & - & 12 & 0 \\
				\hline
				Out. Den. & - & - & SM & $n_c$ & 13*$n_c$ \\
				\hline
				\hline
				\multicolumn{5}{|c|}{\textbf{Total Number of Parameters}} & \textbf{6846+(13*\textbf{$n_c$})}\\ 
				\hline
				
			\end{tabular}
			\begin{tablenotes}
				\item \textit{Segment Input (Seg. Inp.), Dropout (D.O.), Batch Normalization (B.N.), Global Average Pooling (G. Pool), Feature Input (Feat. Inp.), Feature Dense (Feat. Den.), Output Dense (Out. Den.), Softmax (S.M.)}
			\end{tablenotes}
		\end{threeparttable}
	\vspace{-5mm}
	\end{table}

	\section{Experimental Evaluation}
	\subsection{Dataset}
	WESAD dataset is used for the validation of our proposed methodology as it is the only publicly available dataset
that contains wrist-based PPG sensor data for stress and affect detection. Although the dataset contains data for a total of 15 subjects from both chest (RespiBAN) and wrist (Empatica E4) worn sensors, we are only interested in using the wrist-based BVP signal collected through the PPG sensor. The dataset is labeled for 3 types of classes - baseline (neutral), amusement, stress.

	\subsection{Performance Metric}
	As the number of segments for different classes in the dataset is highly imbalanced, only classification accuracy is not appropriate to measure performance. Therefore, the F1 score provides a better measure that balances precision and recall performance. To ensure a fair comparison with our related work in \cite{WESAD}, we use a macro F1 score where each class is given equal importance. The metrics used for evaluation are given below:
	\begin{equation}
	Accuracy = \frac {TP+TN}{TP+FP+TN+FN}
	\end{equation}
	\vspace{-2mm}
	\begin{equation}
	Precision = \frac {TP}{TP+FP}
	\end{equation}
	\vspace{-2mm}
	\begin{equation}
	Recall = \frac {TP}{TP+FN}
	\end{equation}
	\vspace{-2mm}
	\begin{equation}
	Macro~F_1 = \frac {1} {{n_c}} \sum_{i}^{n_c} 2* \frac {Precision_i.Recall_i}{Precision_i+Recall_i}
	\end{equation}
	\vspace{-2mm}
	
	Where TP, TN, FP, FN represents True Positives, True Negatives, False Positives, and False Negatives respectively. The classes are indexed by \textit{i}, and $n_c$ is the number of output classes.
	\subsection{Model Training and Evaluation} 

	We train our normal CNN and H-CNN classifiers with a batch size of 500. The models are trained for 200 epochs with an early stopping mechanism having a patience value of 70. We monitor the validation recall value to select the best model from the epochs. To ensure proper training for the imbalance dataset, we assign class weights to each class using the following formula in Eq. \ref{class_weight}.
	\begin{equation}
	\label{class_weight}
	w_{i} = \frac {1}{N_i} * \frac {N}{n_c}
	\end{equation}

\begin{figure}[t]
		\centering
		\includegraphics[trim={10cm 15.7cm 11cm 5.3cm},clip, width=\linewidth]{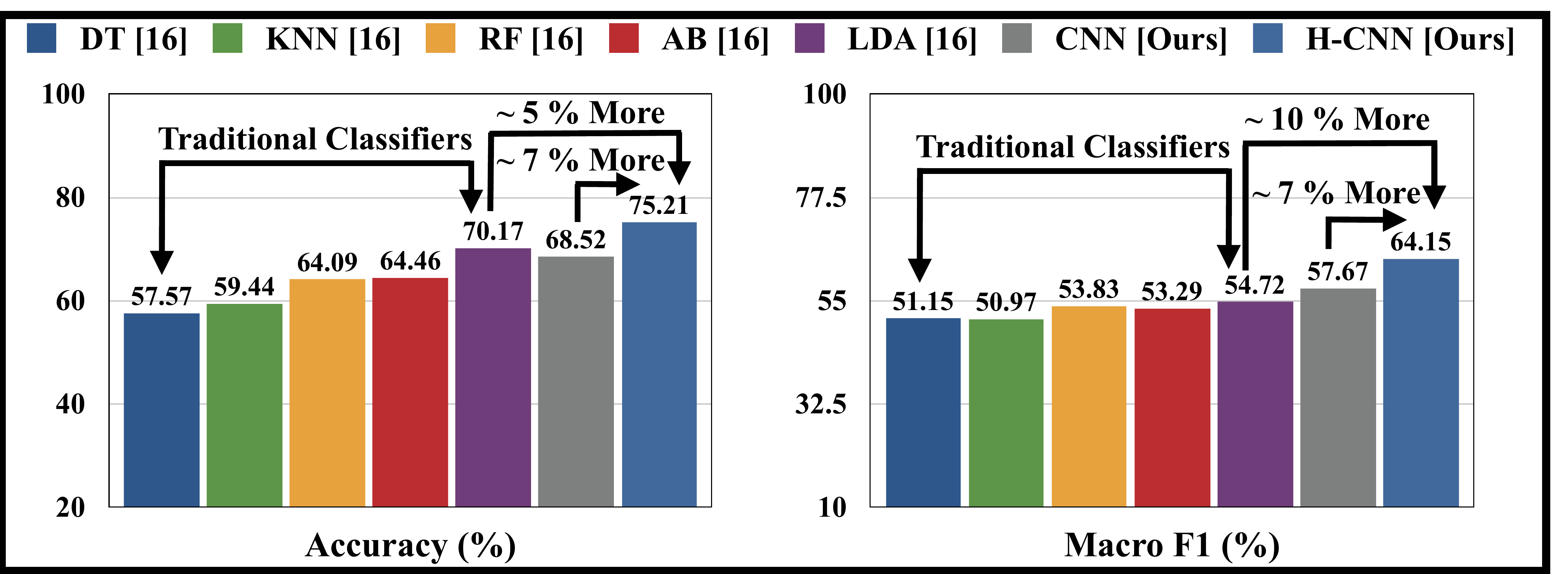}
		\caption{Performance Comparison on 3-Class (Baseline vs Stress vs Amusement) Classification}
		\label{3_Class_Classification}
	\end{figure}

\begin{figure}[t]
		\centering
		\includegraphics[trim={10cm 15.7cm 11cm 5.3cm},clip, width=\linewidth]{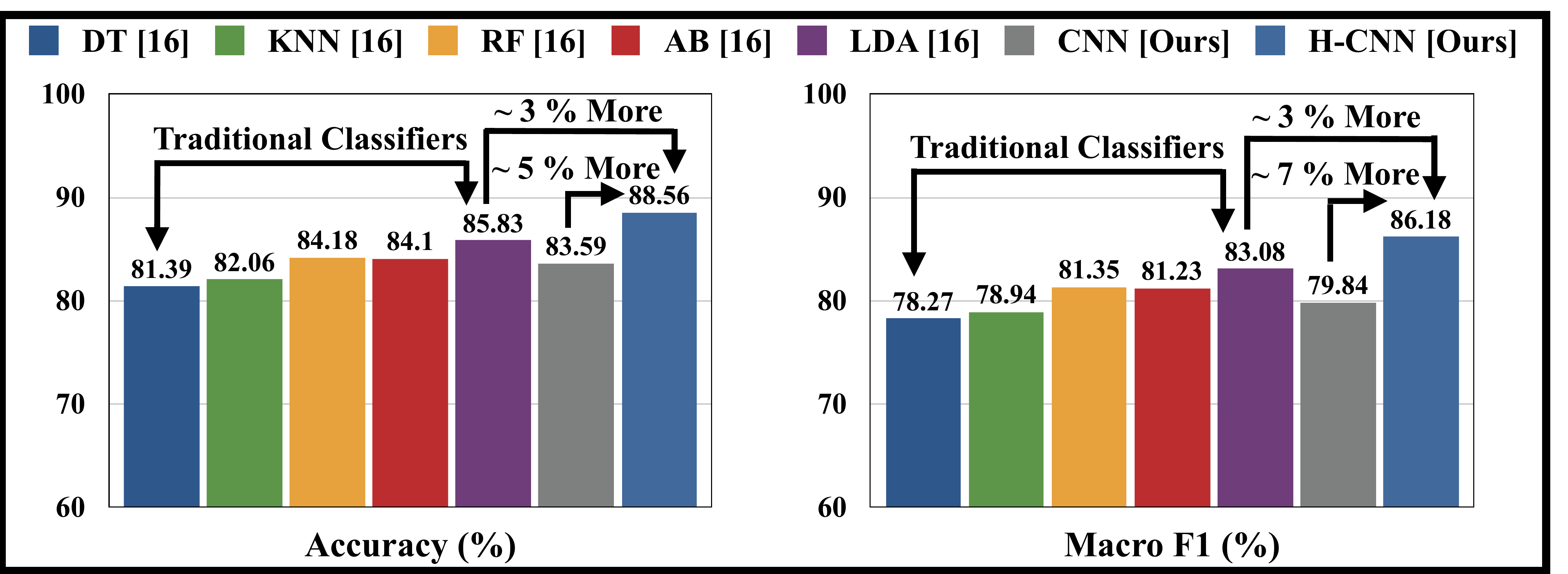}
		\caption{Performance Comparison on 2-Class (Stress vs Non-stress) Classification}
		\vspace{-4mm}
		\label{2_Class_Classification}
	\end{figure}

	Here, $w_{i}$, and $N_i$ represent the class weight and the number of segments belonging to class $i$, respectively. $N$ is the total number of segments from all classes and $n_c$ is the number of output classes.
	The \textit{CategoricalCrossentropy} is used as the loss function. We use the \textit{Adam} optimizer with a learning rate of .001. To demonstrate the generalization property of our trained model and to ensure a fair comparison with the traditional classifiers in \cite{WESAD}, we also perform Leave One Subject Out (LOSO) validation. As shown in Figure \ref{3_Class_Classification}, the Linear Discriminant Analysis (LDA) classifier in \cite{WESAD} outperforms other classical algorithms for 3-class classification with an accuracy of 70.17\% and macro F1 score of 54.72\%. Our normal CNN achieves slightly less accuracy of 68.52\% compared to LDA but outperforms in macro F1 score with 57.67\%. Our H-CNN classifier outperforms both LDA and our normal CNN with an accuracy of 75.21\% and macro F1 score of 64.15\%. Thus, our H-CNN improves the accuracy by $\approx$5\% and $\approx$7\% compared to LDA and normal CNN, respectively. For macro F1 score, our H-CNN shows higher improvement of $\approx$10\% and $\approx$7\% compared to LDA and normal CNN, respectively. For 2-class (Stress vs. Non-stress) classification, baseline and amusement are considered as the non-stress class. As shown in Figure \ref{2_Class_Classification}, for 2-class classification also, our H-CNN improves the accuracy by $\approx$3\% and $\approx$5\% compared to LDA classifier and normal CNN, respectively. Similarly, for macro F1 score, our H-CNN improves the performance by $\approx$3\% and $\approx$7\% compared to LDA and normal CNN, respectively.
	 
	\section{Conclusion}
 This paper proposes a novel hybrid CNN (H-CNN) classifier to detect stress using a wrist-based PPG sensor focusing on consumer-grade wristwatches. Our H-CNN uses both the hand-crafted features and the automatically extracted features by CNN to detect stress using the BVP signal. Evaluation on the benchmark WESAD dataset shows that, for 3-class classification (Baseline vs. Stress vs. Amusement), our proposed H-CNN outperforms traditional classifiers and normal CNN by $\approx$5\% and $\approx$7\% accuracy, and $\approx$10\% and $\approx$7\% macro F1 score, respectively. Also for 2-class classification (Stress vs. Non-stress), our proposed H-CNN outperforms traditional classifiers and normal CNN by $\approx$3\% and $\approx$5\% accuracy, and $\approx$3\% and $\approx$7\% macro F1 score, respectively. To the best of our knowledge, our H-CNN shows the highest performance for both 3-class and 2 -class classification using the BVP signal from the WESAD dataset while performing LOSO validation.
	
	\section{Acknowledgement}
	This work is partially supported by the National Institutes of Health (NIH) grant R41DA049615 and the Graduate Assistance in Areas of National Need (GAANN) award from the United States Department of Education. This paper reflects the views of the authors, not the funding agency.

\end{document}